\documentclass[aps,prb,a4paper,10pt,twocolumn,showpacs,floatfix,superscriptaddress,amsmath,amsfonts,amssymb,preprintnumbers,longbibliography]{revtex4-1}
\usepackage{float}
\usepackage{dcolumn,graphicx,color,booktabs,microtype,afterpage}
\usepackage{amssymb}
\usepackage{amsmath}
\graphicspath{{./}{figure/}}
\usepackage[charter,greekuppercase=italicized]{mathdesign}
\usepackage{sidecap}
\usepackage[mathlines]{lineno}
\renewcommand{\tablename}{Table}
\makeatletter\renewcommand{\fnum@figure}[1]{\figurename~\thefigure.~}\makeatother
\makeatletter\renewcommand{\fnum@table}[1]{\tablename~\thetable.}\makeatother

\newcount\hh \newcount\mm
\hh=\time \divide\hh by 60
\mm=\hh \multiply\mm by 60 \mm=-\mm
\advance\mm by \time
\def\now{\number\hh:\ifnum\mm<10{}0\fi\number\mm}

\usepackage[colorlinks,plainpages=false,linkcolor=blue,urlcolor=blue,citecolor=blue,pdfpagemode=UseNone,pdfstartview=FitBH]{hyperref}

\newcommand{\tcr}[1]{\textcolor{black}{#1}}
\newcommand{\tcb}[1]{\textcolor{black}{#1}}

\newcommand{\tabincell}[2]{\begin{tabular}{@{}#1@{}}#2\end{tabular}}

\begin{document}

\makeatletter\renewcommand{\ps@plain}{%
\def\@evenhead{\hfill\itshape\rightmark}%
\def\@oddhead{\itshape\leftmark\hfill}%
\renewcommand{\@evenfoot}{\hfill\small{--~\thepage~--}\hfill}%
\renewcommand{\@oddfoot}{\hfill\small{--~\thepage~--}\hfill}%
}\makeatother\pagestyle{plain}

\preprint{\textit{Preprint: \today, \now.}} %For internal use only, do not distribute.}}
%\linenumbers

%
%\title{Crystal structure and superconductivity of binary Re$_{1-x}$Mo$_x$ alloys}
%\title{Two domes of superconductivity in binary Re$_{1-x}$Mo$_x$ alloys}
%\title{Emergence of double-dome superconductivity in the binary Re$_{1-x}$Mo$_x$ alloy}
\title{Structure and superconductivity in the binary Re$_{1-x}$Mo$_x$ alloys} % For coherence with the first line of the abstract: plural
\author{T.\ Shang}\email[Corresponding authors:\\]{tian.shang@psi.ch}
\affiliation{Laboratory for Multiscale Materials Experiments, Paul Scherrer Institut, Villigen CH-5232, Switzerland}
\affiliation{Swiss Light Source, Paul Scherrer Institut, Villigen CH-5232, Switzerland}
\affiliation{Institute of Condensed Matter Physics, \'Ecole Polytechnique F\'ed\'erale de Lausanne (EPFL), Lausanne CH-1015, Switzerland.}
%
%\author{M.\ Smidman}\email[Corresponding authors:\\]{msmidman@zju.edu.cn}
%\affiliation{Center for Correlated Matter and Department of Physics, Zhejiang University, Hangzhou 310058, China}
%
\author{D.\ J.\ Gawryluk}\email[Corresponding authors:\\]{dariusz.gawryluk@psi.ch}
\affiliation{Laboratory for Multiscale Materials Experiments, Paul Scherrer Institut, Villigen CH-5232, Switzerland}
\author{J.\ A.\ T.\ Verezhak}
\affiliation{Laboratory for Muon-Spin Spectroscopy, Paul Scherrer Institut, CH-5232 Villigen PSI, Switzerland}
\author{E.\ Pomjakushina}
\affiliation{Laboratory for Multiscale Materials Experiments, Paul Scherrer Institut, Villigen CH-5232, Switzerland}
\author{M.\ Shi}
\affiliation{Swiss Light Source, Paul Scherrer Institut, Villigen CH-5232, Switzerland}
\author{M.\ Medarde}
\affiliation{Laboratory for Multiscale Materials Experiments, Paul Scherrer Institut, Villigen CH-5232, Switzerland}
\author{J.\ Mesot}
\affiliation{Paul Scherrer Institut, CH-5232 Villigen PSI, Switzerland}
\affiliation{Institute of Condensed Matter Physics, \'Ecole Polytechnique F\'ed\'erale de Lausanne (EPFL), Lausanne CH-1015, Switzerland.}
\affiliation{Laboratorium f\"ur Festk\"orperphysik, ETH Z\"urich, CH-8093 Zurich, Switzerland}
\author{T.\ Shiroka}\email[Corresponding authors:\\]{tshiroka@phys.ethz.ch}
\affiliation{Laboratorium f\"ur Festk\"orperphysik, ETH Z\"urich, CH-8093 Zurich, Switzerland}
\affiliation{Paul Scherrer Institut, CH-5232 Villigen PSI, Switzerland}

\begin{abstract}
The binary Re$_{1-x}$Mo$_x$ alloys, known to cover the full range of solid solutions, were successfully synthesized and their crystal structures and physical properties  
investigated via powder x-ray diffraction, electrical resistivity, magnetic susceptibility, and heat capacity. By varying the Re/Mo ratio we explore the full Re$_{1-x}$Mo$_x$ binary phase diagram, in all its four different solid phases: hcp-Mg ($P6_3/mmc$), $\alpha$-Mn ($I\overline{4}3m$), $\beta$-CrFe ($P4_2/mnm$), and bcc-W ($Im\overline{3}m$), of which the second is non-centrosymmetric with the rest being centrosymmetric. All Re$_{1-x}$Mo$_x$ alloys are superconductors, whose critical temperatures exhibit a peculiar phase diagram, characterized by three different superconducting \tcr{regions}. In most alloys the $T_c$ is almost an order of magnitude higher than in pure Re and Mo. Low-temperature electronic specific-heat data evidence a fully-gapped superconducting state, whose enhanced gap magnitude and specific-heat discontinuity suggest a moderately strong electron-phonon coupling across the series. 
Considering that \tcr{several} $\alpha$-Mn-type Re$T$ alloys ($T$ = transition metal) show time-reversal symmetry breaking (TRSB) in the superconducting state, 
while TRS is preserved in the isostructural Mg$_{10}$Ir$_{19}$B$_{16}$ or Nb$_{0.5}$Os$_{0.5}$, the Re$_{1-x}$Mo$_x$ alloys represent another suitable system for studying the interplay of space-inversion, gauge, and time-reversal symmetries in future experiments expected to probe TRSB in the Re$T$ family.        
\end{abstract}

% PACS, the Physics and Astronomy Classification Scheme.
%\pacs{74.20.Fg, 74.25.-q, 75.40.Cx, 67.80.dk, 76.60.Cq}
% 74.20.Fg  BCS theory and its development
% 74.25.−q  Properties of superconductors
% 75.40.Cx  Static properties (order parameter, static susceptibility, heat capacities, critical exponents, etc.)
% 75.10.Jm  Quantized spin models, including quantum spin frustration
% 62.50.-p  High-pressure effects in solids and liquids
% 67.80.dk  Magnetic properties, phases, and NMR
% Other possible PACS
% 76.60.Cq  Chemical and Knight shifts
% 76.60.-k  Nuclear magnetic resonance and relaxation
% 75.10.Pq  Spin chain models
% 75.30.Et  Exchange and superexchange interactions (see also 71.70.-d Level splitting and interactions)
% 75.30.Hx  Magnetic impurity interactions
% 75.45.+j  Macroscopic quantum phenomena in magnetic systems
% 05.10.Cc  Renormalization group methods
% 05.10.Ln  Monte Carlo methods
% 61.50.Ks  Crystallographic aspects of phase transformations; pressure effects (see also 81.30.Hd in materials science)

%\keywords{Unconventional supeconductivity, optimal doping, multiband effects, magnetism, nuclear magnetic resonance}

\maketitle\enlargethispage{3pt}

\vspace{-5pt}
\section{Introduction\label{sec:Introduction}}\enlargethispage{8pt}
Time reversal and spatial inversion are two key symmetries which radically influence electron pairing in the superconducting state. 
Superconductors with a space-inversion center can host either even-parity spin-singlet (e.g., $s$- or $d$-wave) or odd-parity 
spin-triplet (e.g., $p$-wave) pairing states. These strict symmetry requirements, however, are relaxed in non-centrosymmetric superconductors (NCSCs), where the antisymmetric spin-orbit coupling (ASOC) allows, in principle, the occurrence of parity-mixed superconducting states, whose 
mixing degree is related to the strength of the ASOC and to other microscopic parameters.~\cite{Bauer2012,smidman2017}
Because of the mixed pairing,  
NCSCs frequently display \tcr{interesting} properties.
For instance, some NCSCs, such as CePt$_3$Si,\cite{bonalde2005CePt3Si} CeIrSi$_3$,\cite{mukuda2008CeIrSi3}
Li$_2$Pt$_3$B,\cite{yuan2006,nishiyama2007} K$_2$Cr$_3$As$_3$,\cite{K2Cr3As3Pen,K2Cr3As3MuSR} and YBiPt~\cite{Kim2018} exhibit 
line nodes in the gap, while others, such as LaNiC$_2$~\cite{chen2013} and (La,Y)$_2$C$_3$,\cite{kuroiwa2008} 
show multiple-gap superconductivity. The external pressure drives CeIrSi$_3$ into a gapless superconductivity,~\cite{Landaeta2018} while 
a nodal behavior has been 
observed in LaNiC$_2$ and Y$_2$C$_3$.~\cite{Bonalde2011,Landaeta2017,ChenY2C32011}
%Moreover, a strong ASOC lifts the degeneracy of the conduction electrons, 
%resulting in upper critical fields exceeding the Pauli limit (provided the orbital limiting field is sufficiently large) as found in (Ta,Nb)Rh$_2$B$_2$ and 
%CePt$_3$Si.\cite{Carnicom2018,bauer2004} 
Recently, many NCSCs,\cite{Sato2009,Sato2009,Tanaka2010,Chadov2010,Meiner2016, Sun2015, smidman2017,Kim2018,Ali2014} 
in particular YPtBi,\cite{Kim2018} BiPd,~\cite{Sun2015} and PbTaSe$_2$,\cite{Ali2014} 
have been closely investigated as possible models of topological superconductors.

Interestingly, numerous muon-spin relaxation/rotation ($\mu$SR) studies have 
revealed that some NCSCs exhibit also time-reversal symmetry breaking (TRSB), 
concomitant with the onset of superconductivity. Examples include LaNiC$_2$,\cite{Hillier2009} 
La$_7$(Ir,Rh)$_3$,\cite{Barker2015,Singh2018La7Rh3} and \tcr{several} Re-based binary alloys 
Re$T$($T$= transition metal, e.g., Ti, Zr, Nb, Hf).\cite{Singh2014,Singh2017,Singh2018,Shang2018,TianReNb2018} In general, the breaking of time-reversal 
symmetry below $T_c$ and a lack of space-inversion symmetry of the crystal structure are independent events, not required to occur together. For instance, TRS is broken in \tcr{several} $\alpha$-Mn-type Re$T$ 
compounds and in the pure elementary Re,~\cite{TianReNb2018} yet it is preserved in the isostructural Mg$_{10}$Ir$_{19}$B$_{16}$ or Nb$_{0.5}$Os$_{0.5}$,\cite{Acze2010,SinghNbOs} clearly suggesting that TRS breaking is most likely
related to the presence of Re atoms, rather than to a generic lack of space-inversion symmetry. 
Indeed, by converse, the centrosymmetric Sr$_2$RuO$_4$, PrOs$_4$Ge$_{12}$, and LaNiGa$_2$ also exhibit 
a broken TRS in the superconducting state.\cite{Luke1998,aoki2003,Hillier2012} 
%thus making a causal link between time-reversal- and space-inversion symmetry 
%breaking rather unlikely. 

%Despite the plausibility of the above examples, they still refer to different 
%materials (although some of them share similar structures), thus 
%leaving open the cause of TRS breaking in Re$T$ materials 
% and of its possible link to a lack of  space-inversion symmetry.
%To rigorously address the above puzzle 

To further study the TRSB in Re$T$ materials, one should identify 
a system that exhibits both centro- and non-centrosymmetric structures, 
while still preserving its basic stoichiometry. For instance, 
depending on synthesis protocol, Re$_3$W can be either a centro- 
(hcp-Mg-type) or a non-centrosymmetric ($\alpha$-Mn-type) 
superconductor,\cite{Biswas2011} yet neither is found to break 
TRS.\cite{Biswas2012} On the other hand, other superconducting 
Re$T$ compounds, with $T$ = Ti, Zr, Nb, Hf, indeed break TRS, yet mostly adopt the same ($\alpha$-Mn-type) structure. 
%Unlike the above cases,  
%the Re$_{1-x}$Mo$_{x}$ binary alloys discussed here 
%represent the ideal candidate system. 
\tcr{Similar to the Re$_3$W case, the Re$_{1-x}$Mo$_{x}$ binary alloys discussed here represent another candidate system.} For different Re/Mo ratios, 
depending on synthesis protocol, they adopt either centro- or non-centrosymmetric 
structures.\cite{Okamoto1996} 
Although the superconductivity of several Re$_{1-x}$Mo$_{x}$ alloys 
was reported decades ago, only recently the Mo-rich side was 
studied by different techniques.\cite{Roberts1976,Shum1986,Okada2013,Ignatyeva2004} To date, a 
systematic study of the full range of Re$_{1-x}$Mo$_{x}$ 
solid solution is missing. 
In particular, due to synthesis difficulties, its Re-rich side remains 
largely unexplored. Yet, in view of 
the non-centrosymmetric structures adopted, 
it is precisely this part of the phase diagram to be the most interesting one. 

In this paper, based on systematic physical-property measurements, we explore the full superconducting phase diagram of 
the Re$_{1-x}$Mo$_x$ system. To this aim, polycrystalline Re$_{1-x}$Mo$_x$ samples, with 
$0.12 \le$ x $\le 0.75$, were successfully synthesized. 
Although samples with different Re/Mo ratios exhibit different crystal 
structures, they all become superconductors (whose highest $T_c$ 
reaches 12.4\,K). All the relevant superconducting parameters, 
including gap values and symmetries, % here referred to different x values %its symmetry, 
were determined by magnetometry, transport, and specific-heat 
measurements, thus allowing us to present the complete Re$_{1-x}$Mo$_x$ 
superconducting phase diagram.

After briefly describing the experimental methods in Sec.~\ref{sec:details}, 
we present the key results in Sec.~\ref{sec:results}, including those of 
EDX and XRD, electrical resistivity, magnetization, and specific heat.  
Finally, the overall superconducting phase diagram is presented and 
discussed in Sec.~\ref{ssec:Phase}. 

\section{Experimental details\label{sec:details}}\enlargethispage{8pt}
%vspace{3mm}
Polycrystalline Re$_{1-x}$Mo$_x$ (${0.12} \le x \le {0.75}$) alloys were prepared by arc melting 
Re and Mo metals with different stoichiometric ratios in high-purity argon atmosphere. To improve the homogeneity, samples were flipped and remelted several times and, for some of them, the as-cast ingots were annealed at 900$^\circ$C for two weeks. The $\beta$-CrFe phase (e.g, Re$_{0.6}$Mo$_{0.4}$) was obtained by interrupting the heating immediately after the melting of the precursors. 
Hence, all the measurements reported here for %Re$_{0.6}$Mo$_{0.4}$ 
the $\beta$-CrFe phase refer to as-cast samples. % NO "an" 
The extra phases were obtained by further annealing the as-cast samples. The $\alpha$-Mn phase with a non-centrosymmetric crystal structure (Re$_{0.77}$Mo$_{0.23}$) was stabilized by annealing the sample over one week at 1400$^\circ$C in argon or hydrogen atmosphere. 
Unlike the rather malleable Mo-rich alloys, their Re-rich counterparts turned 
out to be extremely hard. In addition, the Re$_{0.77}$Mo$_{0.23}$ alloy resulted fragile after annealing at 1400$^\circ$C. 
Previously, the same arc-melting processes were adopted to cover the whole $x$ range,\cite{Farzadfar2009,Yang2010} 
with the samples being annealed at 1200$^\circ$C and then quenched in water. However, these early attempts 
failed to produce clean $\alpha$-Mn phase.
 
The x-ray powder diffraction (XRD) patterns were measured at room temperature by using a Bruker D8 diffractometer with Cu K$\alpha$ radiation. The atomic ratios of the Re$_{1-x}$Mo$_x$ samples were measured by x-ray fluorescence spectroscopy (XRF) on an AMETEK Orbis Micro-XRF analyzer. The magnetic susceptibility, electrical resistivity, and specific-heat measurements were performed on a 7-T Quantum Design Magnetic Property Measurement System (MPMS-7) and a 9-T Physical Property Measurement System (PPMS-9). 

%\tcr{Description of calculation method.}

\section{\label{sec:results}Key experimental results}\enlargethispage{8pt}% and discussion
\subsection{\label{ssec:structure}Crystal structures}
\begin{figure}[tb]
	\centering
	\includegraphics[width=0.45\textwidth]{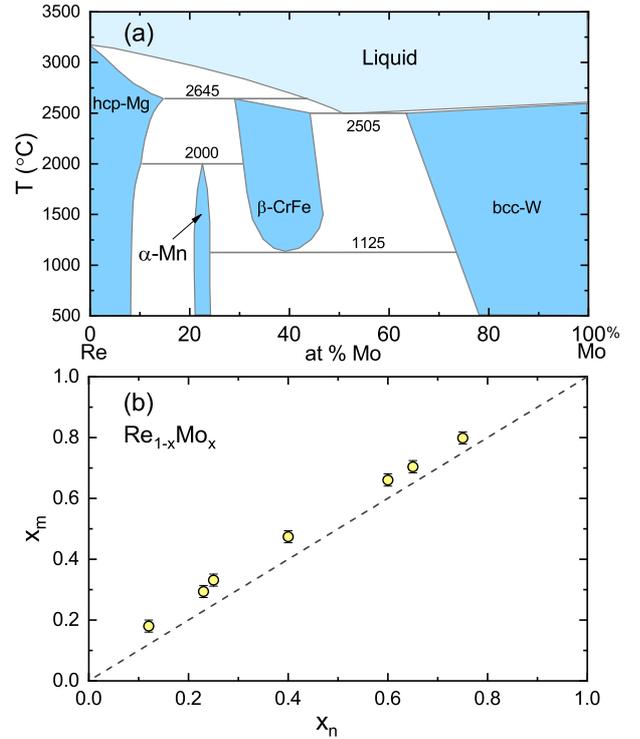}
	\caption{\label{fig:binary}(a) Binary phase diagram of Re$_{1-x}$Mo$_x$ alloys 
	(data adapted from Ref.~\onlinecite{Okamoto1996}). In our case, more than ten 
	samples with different Re/Mo ratios were synthesized. (b) Measured Mo concentration $x_{m}$ vs.\ the nominal $x_{n}$ value. The dashed-line refers to the 
	ideal case where the two values coincide.} 
\end{figure}

As shown in Fig.~\ref{fig:binary}(a), the Re$_{1-x}$Mo$_x$ binary alloys 
exhibit a very rich phase diagram. While pure Mo and Re form body-centered-cubic 
(bcc) and hexagonal-close-packed (hcp) structures, respectively, at 
intermediate Re/Mo ratios two other phases appear, a tetragonal $\beta$-CrFe and a cubic $\alpha$-Mn phase. 
By suitably combining arc melting and annealing processes, we could 
successfully synthesize pure-phase alloys representative of all  
four crystal structures. Although the Re$_{1-x}$Mo$_x$ phase diagram 
has been studied before, most of the obtained samples contained at least 
two different solid phases, thus preventing a systematic study of their 
physical properties.\cite{Farzadfar2009,Yang2010} Conforming to the 
binary phase diagram, the bcc-W and hcp-Mg solid phases were easily
obtained via arc melting of Re and Mo and resulted to be stable at 1400$^\circ$C. 
Two reactions take place during the solidifying process:
\begin{align*}
\textrm{liquid} &\rightarrow \textrm{bcc-W}\, + \,\beta\textrm{-CrFe} 
\qquad \textrm{and} \\ \nonumber
\textrm{liquid + hcp-Mg} &\rightarrow \beta\textrm{-CrFe}.
\end{align*}
In order to synthesize the pure $\beta$-CrFe phase, the first reaction 
was blocked by interrupting the heating imediately after the melting of both the Re and 
Mo metals. As for the $\alpha$-Mn phase, this cannot be synthesized via 
arc melting, since no such phase appears during the liquid-mixture cooling. 
Yet, two other reactions include this phase: 
\begin{align*}
\beta\textrm{-CrFe + hcp-Mg} &\rightarrow \alpha\textrm{-Mn} 
\qquad \qquad \qquad  \textrm{and} \\ \nonumber
\beta\textrm{-CrFe} &\rightarrow \textrm{bcc-W} + \alpha\textrm{-Mn}. 
\end{align*}
Since in both cases solid-state reactions are involved, the pure $\alpha$-Mn 
phase was obtained by annealing the melted Re and Mo metals at 1400$^\circ$C. 

Subsequently, the Re/Mo atomic ratio was determined via EDX on polished samples. 
The estimated Mo (or Re) concentration vs.\ its nominal value is presented 
in Fig.~\ref{fig:binary}(b). For all the samples, the measured Re (Mo) concentration 
was slightly smaller (larger) than the nominal value, reflecting the preferential 
evaporation of Re during the arc melting process. For clarity, since 
typical deviations do not exceed ca.\ 12\%, the nominal concentrations
will be used hereafter.     

%==== figure =============================%
%=== end figure ==========================%
%

\begin{table*}[tbp]
	\centering
	\caption{\label{tab:lattice}Crystal structure, space group, and refined lattice parameters for four representative phases of the Re$_{1-x}$Mo$_x$ binary alloy, 
	with Mo concentrations $x = 0.12$, 0.23, 0.4, and 0.6.} 
	\begin{ruledtabular}	
					 \begin{tabular}{lccccc}
		    Sample        & Re$_{0.88}$Mo$_{0.12}$      &	Re$_{0.77}$Mo$_{0.23}$                &    Re$_{0.6}$Mo$_{0.4}$          &	Re$_{0.4}$Mo$_{0.6}$ \\ \hline	
			Structure     & hexagonal hcp-Mg             &   cubic $\alpha$-Mn                     &    tetragonal $\beta$-CrFe       &  cubic bcc-W \\
			Space group   & $P6_3/mmc$ (No.\,194)       &   $I\overline{4}3m$ (No.\,217)          &    $P4_2/mnm$ (No.\,136)         &  $Im\overline{3}m$(No.\,229)\\
			$a$ (\AA{})      & 2.76798(2)                  &   9.58476(3)                            &    9.58514(4)                    &  3.12627(10)\\
			$b$ (\AA{})      & 2.76798(2)                  &   9.58476(3)                            &    9.58514(4)                    &  3.12627(10)\\
 			$c$ (\AA{})      & 4.48728(5)                  &   9.58476(3)                            &    4.97891(2)                    &  3.12627(10)\\
			$V_\mathrm{cell}$ (\AA{}$^3$)  & 29.7743(5)  &   880.529(6)                            &    457.437(4)                    &  30.5549(17)\\
		\end{tabular}
			\end{ruledtabular}
\end{table*}

\begin{table}	
	\centering
	\caption{\label{tab:coordinate}Refined atomic coordinates and site occupancy factor (SOF) of four different crystal structures of Re$_{1-x}$Mo$_x$ alloys with $x = 0.12$, 0.23, 0.4, and 0.6.}
	\begin{ruledtabular}	
	     \vspace{8pt}
	    Re$_{0.88}$Mo$_{0.12}$\\
		\begin{tabular}{lccccc}
			\textrm{Atom}&
			\textrm{Wyckoff}&
			\textrm{$x$}&
			\textrm{$y$}&
			\textrm{$z$}&
			\textrm{SOF}\\
			\colrule %\\
			Re1           & 2$c$  & 0.33330  & 0.66670  & 0.25000 	& 0.88 \rule{0pt}{2.6ex} \\
			Mo1           & 2$c$  & 0.33330  & 0.66670  & 0.25000 	& 0.12\\
		\end{tabular}
		\\ \vspace{10pt}
		
		Re$_{0.77}$Mo$_{0.23}$\\
		\begin{tabular}{lcccccc}
			\textrm{Atom}&
			\textrm{Wyckoff}&
			\textrm{$x$}&
			\textrm{$y$}&
			\textrm{$z$}&
         	\textrm{SOF}\\ 
            \colrule %\\
			Mo1            & 8$c$   & 0.3260(4)  & 0.3260(4)   & 0.3260(4)  & 1  \rule{0pt}{2.6ex} \\
			Mo2            & 24$g$  & 0.3593(2)  & 0.3593(2)   & 0.0409(3)  & 0.30(1)         \\
			Re1            & 2$a$   & 0          & 0           & 0          & 1               \\
			Re2            & 24$g$  & 0.3593(2)  & 0.3593(2)   & 0.0409(3)  & 0.70(1)         \\
			Re3            & 24$g$  & 0.0911(2)  & 0.0911(2)   & 0.2839(2)  & 1               \\ 
		\end{tabular} 
		\\ \vspace{10pt}
		
		Re$_{0.6}$Mo$_{0.4}$\\
		\begin{tabular}{lccccl}
			\textrm{Atom}&
			\textrm{Wyckoff}&
			\textrm{$x$}&
			\textrm{$y$}&
			\textrm{$z$}&
			\textrm{SOF}\\
			\colrule %\\
			Re1            & 2$a$   & 0            & 0            & 0          &0.69(1) \rule{0pt}{2.6ex} \\
			Re2            & 4$f$   & 0.3994(2)    & 0.3994(2)    & 0          &0.22(1) \\
			Re3            & 8$i$   & 0.4650(1)    & 0.1303(2)    & 0          &0.52(1) \\
			Re4            & 8$i$   & 0.7418(2)    & 0.0655(2)    & 0          &0.75(1)  \\
			Re5            & 8$j$   & 0.1834(1)    & 0.1834(2)    & 0.25       &0.60(1) \\
		    Mo1            & 2$a$   & 0            & 0            & 0          &0.31(1)  \\
			Mo2            & 4$f$   & 0.3994(2)    & 0.3994(2)    & 0          &0.78(1) \\
			Mo3            & 8$i$   & 0.4650(1)    & 0.1303(2)    & 0          &0.48(1) \\
			Mo4            & 8$i$   & 0.7418(2)    & 0.0655(2)    & 0          &0.25(1)  \\
			Mo5            & 8$j$   & 0.1834(1)    & 0.1834(2)    & 0.25       &0.40(1) \\
			\end{tabular}
		\\ \vspace{10pt}
		
     	Re$_{0.4}$Mo$_{0.6}$\\
		\begin{tabular}{lccccc}
			\textrm{Atom}&
			\textrm{Wyckoff}&
			\textrm{$x$}&
			\textrm{$y$}&
			\textrm{$z$}&
			\textrm{SOF}\\
			\colrule %\\
			Re1           & 2$a$    & 0       & 0        & 0    & 0.4 \rule{0pt}{2.6ex} \\
			Mo1           & 2$a$    & 0       & 0        & 0    & 0.6\\
		\end{tabular}
	\end{ruledtabular}
\end{table}

%
%==== figure =============================%
\begin{figure}[th]
	\centering
	\includegraphics[width=0.4\textwidth,angle=0]{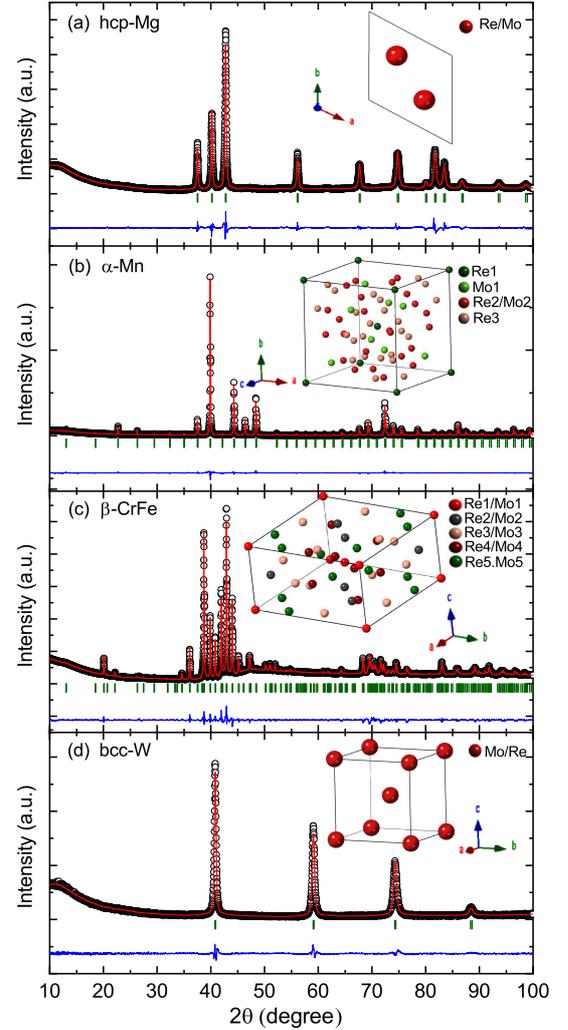}
	\vspace{-2ex}%
	\caption{\label{fig:XRD} Room-temperature XRD patterns and refinements
		for Re$_{0.88}$\-Mo$_{0.12}$ (a),  Re$_{0.77}$Mo$_{0.23}$ (b),  Re$_{0.6}$\-Mo$_{0.4}$ (c), and Re$_{0.4}$\-Mo$_{0.6}$ (d). The black open circles and the solid red lines represent experimental patterns and Rietveld-refinement profiles, respectively. The blue lines show the residuals, i.e., 
		the difference between calculated and experimental data. 
		The vertical green bars mark the calculated Bragg-peak positions.
		In each case, the (unit-cell) crystal structures are shown in the insets.}
\end{figure}
%=== end figure ==========================%

Figure~\ref{fig:XRD} shows four representative XRD patterns of polycrystalline 
Re$_{1-x}$Mo$_{x}$ samples, with the other samples exhibiting similar 
diffractograms 
(not shown here). All the XRD patterns were analyzed using the FullProf Rietveld-analysis suite.\cite{Carvajal1993}
No obvious impurity phases could be detected, indicating the high quality 
of the synthesized samples. As shown in Fig.~\ref{fig:binary}(a), on the Re-rich side, Re$_{1-x}$Mo$_{x}$ alloys 
adopt a hexagonal hcp-Mg-type structure. According to our XRD refinements 
in Fig.~\ref{fig:XRD}(a), all the samples with $0.12 \leq x \leq 0.25$ 
exhibit hcp-Mg-type structures [see inset in Fig.~\ref{fig:XRD}(a)]. 
Following the binary phase diagram, for $x = 0.23$ 
besides the hexagonal structure, a cubic $\alpha$-Mn-type structure can 
also be stabilized by sample annealing at 1400$^\circ$C [see 
Fig.~\ref{fig:XRD}(b)]. Such cubic phase is the same as that adopted by 
other Re$T$ compounds, where TRSB and unconventional superconductivity 
have been frequently observed (see Sec.~\ref{sec:Introduction}). 
However, unlike in other Re$T$ binary alloys, in Re$_{1-x}$Mo$_{x}$ 
the $\alpha$-Mn-type phase region is extremely narrow [see Fig.~\ref{fig:binary}(a)], i.e., $0.21 \le x \le 0.25$. 
To exclude the hcp-Mg phase, the Re concentration was kept fixed at 
$(1-x) = 0.77$, i.e., at the center of the $\alpha$-Mn-phase region. By 
further increasing Mo concentration, the tetragonal $\beta$-CrFe-type 
solid phase was synthesized %can be synthesized 
via quenching. Its XRD refinement 
and crystal structure are presented in Fig.~\ref{fig:XRD}(c). There are 
five different Re sites in the unit cell. Further on, in the Mo-rich side, 
the Re$_{1-x}$Mo$_{x}$ alloys adopt a cubic bcc-W-type crystal structure. 
As an example, in Fig.~\ref{fig:XRD}(d) we show the XRD refinement of the 
$x = 0.4$ sample, along with its crystal structure (see inset). The 
crystal-structure information and atomic coordinates for all the four 
different solid phases are reported in Tables~\ref{tab:lattice} 
and ~\ref{tab:coordinate}, respectively.

\subsection{\label{ssec:rho}Electrical resistivity}
%
%==== figure =============================%
\begin{figure}[th]
	\centering
	\includegraphics[width=0.48\textwidth,angle=0]{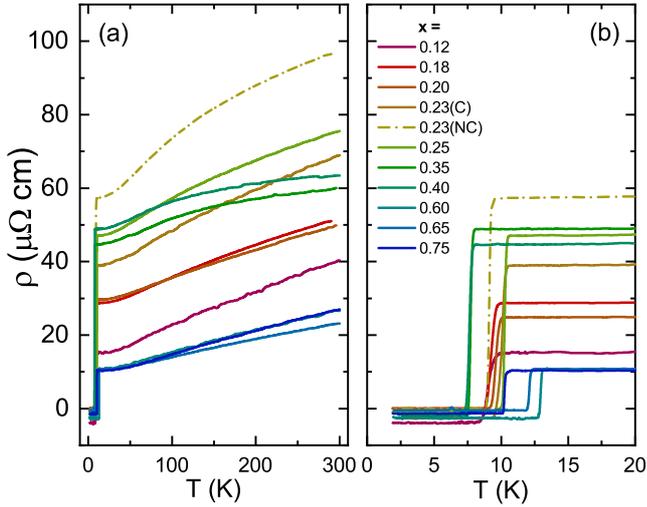}
	\vspace{-2ex}%
	\caption{\label{fig:Rho}(a) Temperature dependence of the electrical 
	resistivity for Re$_{1-x}$Mo$_x$ ($0.12 \le x \le 0.75$). 
	(b) Enlarged low-temperature data region, highlighting the 
	superconducting transitions. For $x = 0.23$, the electrical resistivity 
	of both centro- (C) and non-centro\-sym\-met\-ric (NC) specimens (dashed-dotted line) are 
	shown.}
\end{figure}
\begin{figure}[th]
	\centering
	\includegraphics[width=0.48\textwidth,angle=0]{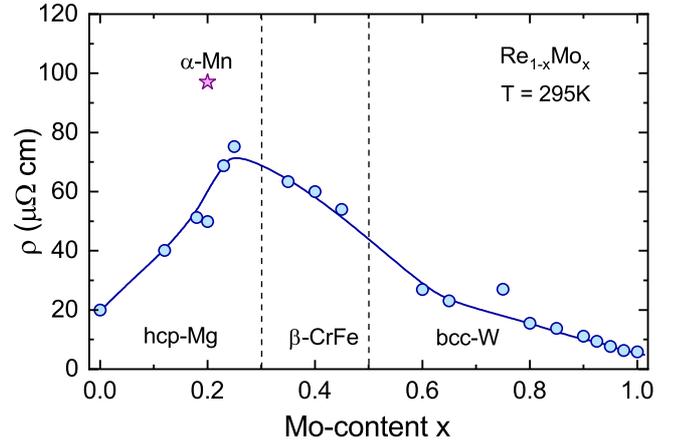}
	\vspace{-2ex}%
	\caption{\label{fig:Rho_RT} Room-temperature electrical resistivity of Re$_{1-x}$Mo$_x$ vs.\ Mo concentration $x$. The star represents the resistivity of Re$_{0.77}$Mo$_{0.23}$ with non-centrosymmetric $\alpha$-Mn-type structure, here 25\% higher than that of its centrosymmetric hcp-Mg counterpart. The dashed lines 
	indicate the solid-phase boundaries. Data for pure Re and  Re$_{1-x}$Mo$_x$ ($0.8 \le x \le 1$)  were taken from Ref.~\onlinecite{Naor2010,Holmwood1965,Sundar2015NJP}.}
\end{figure}

%=== end figure ==========================%
%
The temperature-dependent electrical resistivity $\rho(T)$ of Re$_{1-x}$Mo$_x$ 
($0.12 \le x \le 0.75$) was measured from room temperature down to 2\,K. As shown in Fig.~\ref{fig:Rho}(a), all samples exhibit metallic behavior down to the lowest 
temperature. Apart from the superconducting transition, no anomaly associated with 
structural, magnetic, or charge-density-wave transitions could be detected. 
The room-temperature (295\,K) electrical resistivities vs.\ Mo 
concentration $x$ are summarized in Fig.~\ref{fig:Rho_RT}. Upon increasing 
Mo content, the electrical resistivity also increases, before reaching 
the first phase boundary between hcp-Mg and $\beta$-CrFe phases. We 
note that, for $x = 0.23$, a non-centrosymmetric $\alpha$-Mn-type 
sample (shown by a star in Fig.~\ref{fig:Rho_RT}) exhibits a
25\% larger electrical resistivity compared to its centrosymmetric 
hcp-Mg-type counterpart. As $x$ is further increased beyond 0.3, 
the Re$_{1-x}$Mo$_x$ alloys adopt the $\beta$-CrFe phase and the 
electrical resistivity starts to decrease. On the Mo-rich side, i.e., 
in the bcc-W phase, electrical resistivity is much smaller than on the 
Re-rich side, thus demonstrating the better metallicity of this phase. 
Overall, these results clearly indicate a close relationship between the
crystal structure and the electronic properties in 
the binary Re$_{1-x}$Mo$_x$ alloys. 

The electrical resistivity in the low-temperature region is plotted in 
Fig.~\ref{fig:Rho}(b). Despite the varying $T_c$s across the 
Re$_{1-x}$Mo$_x$ series, all the samples exhibit a superconducting transition at low temperature. 
The $T_c$ values vary nonmonotonically  
with  Re (or Mo) concentration, with the maximum $T_c = 13$\,K being 
achieved for $x = 0.6$. $T_c$ values resulting from electrical-resistivity 
data are summarized in Fig.~\ref{fig:phasediagram} (see below). Finally, 
the centrosymmetric $x = 0.23$ specimen shows a $\sim1$ K higher 
$T_c$ than the non-centrosymmetric one. 

%==== figure =============================%
%=== end figure ==========================%

\subsection{\label{ssec:sus}Magnetic susceptibility}
%
%==== figure =============================%
\begin{figure}[th]
	\centering
	\includegraphics[width=0.48\textwidth,angle=0]{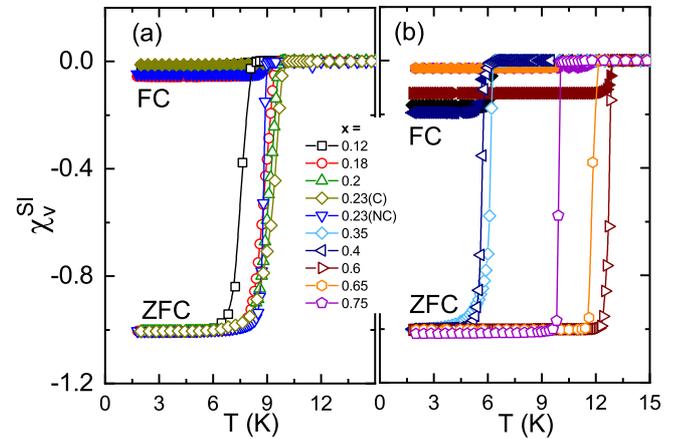}
	\vspace{-2ex}%
	\caption{\label{fig:Chi}Temperature dependence of the magnetic susceptibility 
	$\chi(T)$ of Re$_{1-x}$Mo$_x$ ($0.12 \le x \le 0.75$). Data were collected 
	in an applied field of 1\,mT and a temperature range of 1.8 to 15\,K. The magnetic susceptibilities were corrected by using the demagnetization factors obtained from the field-dependent magnetization at base temperature.}
\end{figure}
%=== end figure ==========================%
%

The bulk superconductivity of Re$_{1-x}$Mo$_x$ alloys was further confirmed by magnetic susceptibility measurements. The temperature dependence of the magnetic 
susceptibility $\chi(T)$ was measured using both field-cooled (FC) and 
zero-field-cooled (ZFC) protocols in an applied field of 1\,mT. As shown 
in Fig.~\ref{fig:Chi}(a)-(b), all the samples show superconducting 
transitions with differing $T_c$s, consistent with the 
respective values determined from electrical resistivity [see 
Fig.~\ref{fig:Rho}]. The splitting of the FC- and ZFC susceptibilities 
is a typical feature of type-II superconductors, where the magnetic-field flux is pinned once the material is cooled in 
an applied field. Since $\chi_\mathrm{v}^\mathrm{SI} \sim 1$, the 
ZFC-susceptibility data indicate (almost ideal) bulk superconductivity 
below $T_c$. %In the whole Re$_{1-x}$Mo$_x$ series, the $x = 0.4$ and 0.6 samples show the most significant FC curves, suggesting the strongest flux-pinning effect in these two cases. 
  
\subsection{\label{ssec:Cp} Specific heat}
%%
%==== figure =============================%
\begin{figure}[th]
	\centering
	\includegraphics[width=0.48\textwidth,angle=0]{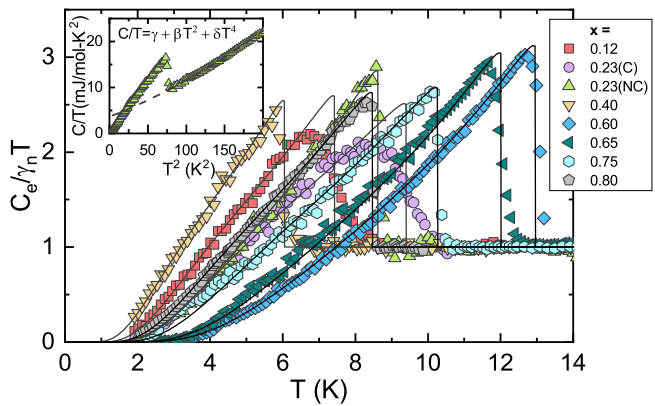}
	\vspace{-2ex}%
	\caption{\label{fig:Cp} Temperature dependence of the normalized electronic specific heat $C_\mathrm{e}/\gamma_n T$ for Re$_{1-x}$Mo$_x$ with $x = 0.12$, 0.20, 0.23 (C and NC), 0.40, 0.6, 0.75 and 0.80. Inset: specific heat $C/T$ as a function of $T^2$ for the  non-centrosymmetric Re$_{0.77}$Mo$_{0.23}$. The dashed-line in the inset is a fit to $C/T = \gamma + \beta T^2 + \delta T^4$, while the solid lines in the main panel are fits to a fully-gapped $s$-wave model. Data for $x$ = 0.80 were taken from Ref.~\onlinecite{Sundar2015NJP}.}  % Data is the plural of "Datum", hence it should be "were" and not "was"
\end{figure}
%=== end figure ==========================%

%==== Table =============================%
\begin{table*}[!bht]
	\centering
	\caption{Superconducting transition temperatures $T_c$, specific-heat fit parameters, estimated Debye temperatures, density of states, and electron-phonon coupling constants for the Re$_{1-x}$Mo$_x$ series.\label{tab:Cp}}  
	\begin{ruledtabular}
		\begin{tabular}{lccccccccc}
			\;\;$x$&
			%$T_c$ (K)&
			\tabincell{c}{$T^{Cp}_c$(K)}&
			%\tabincell{c}{$T^{\rho}_c$(K)}&
			%\tabincell{c}{$T^{\chi}_c$(K)}&
			\tabincell{c}{$\gamma_\mathrm{n}$(mJ/mol-K$^2$)}&
			\tabincell{c}{$\Theta_\mathrm{D}$(K)}&
			$\Delta C/(\gamma_\mathrm{n}T_c)$&
			\tabincell{c}{$\Delta/(k_\mathrm{B}T_c)$}& %	$\gamma_\mathrm{n}$(mJ/mol-K$^2$)&
			%$\Delta$(k$_\mathrm{B}$$T_c$)&
			$\lambda_\mathrm{ep}$ &	
			\tabincell{c}{$N(\epsilon_\mathrm{F}$)/(eV-f.u.)}&
			\tabincell{c}{$N_\mathrm{band}(\epsilon_\mathrm{F}$)/(eV-f.u.)}&
			$m^*$/$m_e$\\
			%\textrm{x}&
			%\textrm{$T_c$(K)}&
			%\textrm{$\gamma_\mathrm{n}$(mJ/mol-K$^2$)}&
			%\textrm{$\Theta_\mathrm{D}$(K)}&
			%\textrm{$\Delta C/\gamma_\mathrm{n}T_c$}&
			%\textrm{$\Delta$(k$_\mathrm{B}$$T_c$)}&
			%\textrm{N($\epsilon_\mathrm{F}$)}&
			%\textrm{$\lambda_\mathrm{ef}$}\\
			%\colrule\\
			\hline
			0.00\footnotemark[1]    &1.69   &2.30   &415.0   &1.30  &1.71  &0.46  &0.33   & 0.27  &1.46 \rule{0pt}{2.6ex} \\
			0.12                    &7.45   &3.80   &402.9   &1.18  &1.85  &0.66  &1.61   & 0.97  &1.66\\
			0.20                    &9.02   &3.77   &363.7   &0.98  &1.80  &0.72  &1.60   & 0.93  &1.72\\
			0.23\footnotemark[2]    &9.43   &3.53   &333.0   &1.07  &1.80  &0.77  &1.50   & 0.85  &1.77\\
			0.23\footnotemark[3]    &8.65   &3.66   &311.4   &1.90  &2.00  &0.76  &1.55   & 0.88  &1.76\\
			0.35                    &6.30   &3.20   &391.1   &1.51  &1.90  &0.63  &1.36   & 0.83  &1.63\\	
			0.40                    &6.07   &3.01   &520.0   &1.44  &1.82  &0.58  &1.28   & 0.81  &1.58\\
			0.45                    &6.60   &3.45   &397.2   &1.59  &1.85  &0.64  &1.46   & 0.89  &1.64\\
			0.60                    &13.00  &4.05   &341.8   &2.00  &2.14  &0.87  &1.72   & 0.92  &1.87\\
		    0.65                    &12.05  &3.89   &397.3   &1.96  &2.10  &0.79  &1.65   & 0.92  &1.79\\
			0.75                    &10.30  &3.69   &480.6   &1.64  &1.90  &0.69  &1.57   & 0.93  &1.69\\
			0.80\footnotemark[1]    &8.50   &3.65   &420.7   &1.51  &1.87  &0.68  &1.55   & 0.92  &1.68\\
			0.85\footnotemark[1]    &6.74   &3.41   &436.2   &1.50  &1.85  &0.62  &1.45   & 0.89  &1.62\\
			0.90\footnotemark[1]    &3.02   &2.39   &429.9   &1.35  &1.80  &0.51  &1.01   & 0.66  &1.51\\
			1.00\footnotemark[1]    &0.92   &1.83   &460.0   &1.25  &1.70  &0.41  &0.28   & 0.20  &1.41\\
		\end{tabular}
		\footnotetext[1]{Data from Refs.~\onlinecite{Rorer1965,Heiniger1966,Smith1970,Sundar2015,Sundar2015NJP}.}
		\footnotetext[2]{centrosymmetric}
		\footnotetext[3]{non-centrosymmetric}	
	\end{ruledtabular}
\end{table*}
%=== end table ==========================%

In the superconducting state, specific-heat data offer valuable 
insight into the superconducting properties, including the gap 
value and its symmetry. Hence, in the Re$_{1-x}$Mo$_x$ series, we 
performed systematic zero-field specific-heat measurements down to 2\,K. 
As shown in Fig.~\ref{fig:Cp}, in all cases a clear specific-heat jump 
indicates a bulk superconducting transition. 
Again, the $T_c$ values determined by using the equal-entropy method are 
summarized in the superconducting phase diagram below. The electronic specific heat $C_\mathrm{e}$/$T$ was obtained by subtracting the phonon contribution from the experimental data. An example is shown in the inset of Fig.~\ref{fig:Cp}, where the normal-state heat capacity of Re$_{0.77}$Mo$_{0.23}$ (NC) is fitted to $C = \gamma_\mathrm{n} T +\beta T^3 + \delta T^5$, where $\gamma_\mathrm{n}T$ is the normal-state-electronic contribution to heat capacity and $\beta T^3 + \delta T^5$ is the phonon contribution to heat capacity. The specific heat of all the samples 
was fitted using the same formula and the derived $\beta$ and $\delta$ values (as listed in Table~\ref{tab:Cp}) were then used to account for the phonon contribution in the superconducting state. The extrapolation 
to a zero intercept of the $C/T$ data close to 0\,K (in the 
inset of Fig.~\ref{fig:Cp}) indicates a fully superconducting 
volume fraction and a good sample quality. The Debye temperature $\Theta_\mathrm{D}$ can be calculated from $\beta$ values, by using $\Theta_\mathrm{D}$= (12$\pi^4$\,$Rn$/5$\beta$)$^{1/3}$, where $R = 8.314$\,JK$^{-1}$mol$^{-1}$ is the molar gas constant and $n = 1$ is the number of atoms per formula unit. The $\gamma_\mathrm{n}$ and $\Theta_\mathrm{D}$ values of Re$_{1-x}$Mo$_x$ alloys are summarized in Table~\ref{tab:Cp}. The density of states (DOS) at the Fermi level $N(\epsilon_\mathrm{F})$ was evaluated from the expression $N(\epsilon_\mathrm{F})$ = $\pi^2$$k_\mathrm{B}^2$/3$\gamma_\mathrm{n}$, 
where $k_\mathrm{B}$ and $\gamma_\mathrm{n}$ are the Boltzmann constant and the normal-state electronic-specific-heat coefficient.\cite{kittel2005} The electron-phonon coupling constant $\lambda_\mathrm{ep}$, a measure of the attractive interaction between electrons and phonons, can be further estimated from the $\Theta_\mathrm{D}$ and $T_c$ values by applying the semi-empirical McMillan formula:\cite{McMillan1968}
\begin{equation}
\lambda_\mathrm{ep}=\frac{1.04+\mu^*\,\mathrm{ln}(\Theta_\mathrm{D}/1.45\,T_c)}{(1-0.62\,\mu^*)\mathrm{ln}(\Theta_\mathrm{D}/1.45\,T_c)-1.04}.
\end{equation}
The Coulomb pseudopotential $\mu^*$, usually lying in the 0.09--0.18 range, 
is here fixed to 0.13, a commonly used value for the transition metals. As listed 
in Table~\ref{tab:Cp}, the enhanced electron-phonon coupling $\lambda_\mathrm{ep}$ of 
Re$_{1-x}$Mo$_x$ (compared with pure Re or Mo) implies a moderate coupling strength 
of electrons in the superconducting state. Similar $\lambda_\mathrm{ep}$ values 
have been found also in other non-centrosymmetric Re$T$ compounds.\cite{Karki2011,Lue2013ReTi,Khan2016,Singh2016ReHf} 
Finally, the band-structure density of states (DOS) $N_\mathrm{band}(\epsilon_\mathrm{F})$ and the effective mass of quasiparticles $m^*$ can be estimated from the relations $N_\mathrm{band}(\epsilon_\mathrm{F}) = N(\epsilon_\mathrm{F})/(1 + \lambda_\mathrm{ep}$) and $m^* = m_\mathrm{band}(1 + \lambda_\mathrm{ep}$).\cite{kittel2005} Here we take $m_\mathrm{band}$ = $m_e$, where $m_e$ is the mass of free electrons. The calculated values for the Re$_{1-x}$Mo$_x$ alloys are summarized in Table~\ref{tab:Cp}. For the  non-centrosymmetric Re$_\mathrm{0.77}$Mo$_\mathrm{0.23}$, the DOS is comparable 
to the values resulting from band-structure calculations on the similar 
Re$T$ ($T$ = Zr, Nb, Ti) compounds.\cite{Khan2016,winiarski2014} 

Below we discuss the Re$_{1-x}$Mo$_x$ superconducting properties based 
on specific-heat data. The derived electronic specific heats divided by 
the normal-state electronic specific-heat coefficients, i.e., 
$C_\mathrm{e}$/$\gamma_\mathrm{n}$$T$, are shown in the main panel of 
Fig.~\ref{fig:Cp} as a function of temperature. The temperature-dependent superconducting-phase contribution to the entropy can be calculated by using 
the following equation:
\begin{equation}
S = \frac{6\gamma_\mathrm{n}}{\pi^2} \frac{\Delta(0)}{k_\mathrm{B}}  \int^{\infty}_0 [f\mathrm{ln}f+(1-f)\mathrm{ln}(1-f)]\,\mathrm{d}x,
\end{equation}
where $f = (1+e^{E/k_\mathrm{B}T})^{-1}$ is the Fermi function, $\Delta(0)$ 
is the SC gap value at 0\,K, and $E(\epsilon) = \sqrt{\epsilon^2 + \Delta^2(T)}$ 
\tcr{is the excitation energy of quasiparticles (i.e., their dispersion), with 
$\epsilon$ the electron energies measured relative to the chemical potential 
(Fermi energy).}~\cite{Padamsee1973,tinkham1996}
%
%the electron energies in the normal state measured relative to the Fermi energy.  
%$E(\epsilon) = \sqrt{\epsilon^2 + \Delta^2(T)}$, 
%and $E(\epsilon) = \sqrt{\epsilon^2 + \Delta^2(T)}$, 
%\tcr{$\epsilon = x \Delta(0)$, is the quasiparticle energy spectrum}.~\cite{Padamsee1973,tinkham1996}
%
%Here $\epsilon = x \Delta(0)$ is the energy of the normal-state electrons 
%relative to the Fermi energy, and 
Here $\Delta(T) = \Delta(0) \mathrm{tanh} \{ 1.82[1.018(T_\mathrm{c}/T-1)]^{0.51}\}$.\cite{ Carrington2003} Once the entropy is known, the temperature-dependent electronic specific heat in the superconducting state 
can be calculated from $C_\mathrm{e} =T \frac{dS}{dT}$. The solid lines 
in Fig.~\ref{fig:Cp} represent fits with the above model and a single 
isotropic gap. The derived superconducting gap values $\Delta/(k_\mathrm{B}T_c)$ are 
summarized in Table~\ref{tab:Cp}. Except for the pure Re or Mo, the gap 
values of Re$_{1-x}$Mo$_x$ alloys are slightly higher than the weak-copuling BCS value 1.763, thus indicating moderately-coupled superconducting pairs in 
Re$_{1-x}$Mo$_x$. The specific-heat discontinuities at $T_c$, i.e., 
$\Delta C/\gamma_\mathrm{n}$$T_{c}$, are also summarized in Table~\ref{tab:Cp}. 
As shown in Fig.~\ref{fig:Cp}, Re$_{1-x}$Mo$_x$ alloys with $x$ = 0.23 (NC), 
0.40, 0.60, and 0.75, show a larger specific-heat discontinuity than the 
conventional BCS value of 1.43, whereas for $x = 0.12$, 0.20, and 0.23(C), the analogous  
discontinuity is reduced, most likely reflecting a broadening 
of the superconducting transition.  

%\subsection{\label{ssec:DFT}DFT calculation results}
%\tcr{Include here some DFT and band-structure calculations}

\section{\label{ssec:Phase}%R\lowercase{e}$_{1-x}$M\lowercase{o}$_x$ 
Phase diagram and discussion}
\begin{figure}[th]
	\centering
	\includegraphics[width=0.48\textwidth,angle=0]{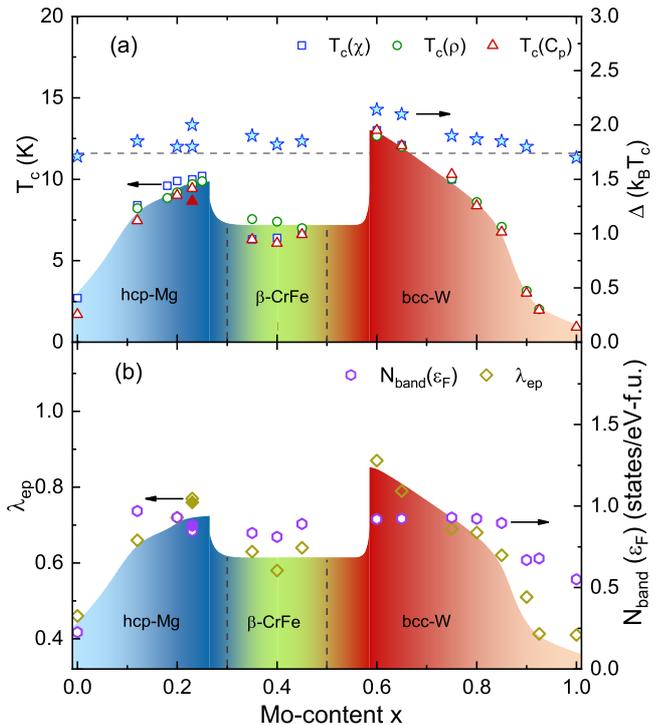}
	\vspace{-2ex}%
	\caption{\label{fig:phasediagram}Superconducting phase diagram of the Re$_{1-x}$Mo$_x$ alloys. (a) Transition temperatures $T_c$, as determined from electrical resistivity, magnetic susceptibility, and specific-heat data, as well as superconducting gap values ($\bigstar$, in $k_\mathrm{B}T_c$ units), as determined from fits to the zero-field specific heat vs.\ Mo content. (b) The calculated band-structure density of states $N_\mathrm{band}(\epsilon_\mathrm{F}$) and the electron-phonon coupling constant $\lambda_\mathrm{ep}$ for Re$_{1-x}$Mo$_x$. The solid symbols refer to the non-centrosymmetric Re$_{0.77}$Mo$_{0.23}$ case. Data for Re$_{1-x}$Mo$_x$ ($0.8 \le x \le 1$)  were adopted from Ref.~\onlinecite{Sundar2015NJP}. Colors highlight the correlation between SC properties and crystal structure.}
\end{figure}

Based on the above experimental data, in Fig.~\ref{fig:phasediagram}, 
we present the superconducting phase diagram of the Re$_{1-x}$Mo$_x$ alloys 
as a function of the Re/Mo concentration. According to the binary phase 
diagram (Fig.~\ref{fig:binary}) and to XRD refinements (Fig.~\ref{fig:XRD}), 
a change in the relative Re/Mo content induces up to four different solid 
phases: %can be stabilized, i.e., 
hcp-Mg ($P6_3/mmc$), $\alpha$-Mn ($I\overline{4}3m$), $\beta$-CrFe ($P4_2/mnm$), 
and bcc-W ($Im\overline{3}m$). 
%The superconducting properties of the Re$_{1-x}$Mo$_x$ alloys seem to reflect closely 
%the respective crystal structures. 

As the Mo concentration increases, the $T_{c}$ 
varies nonmonotonically, giving rise to three distinct superconducting \tcr{regions} 
(see Fig.~\ref{fig:phasediagram}). The first superconducting \tcr{region}, with highest 
$T_c = 9.43$\,K, is achieved on the Re-rich part for $x = 0.23$ 
and corresponds to samples adopting an hcp-Mg-type structure. Note that this $T_c$ is ca.\ 1\,K 
lower than that of the corresponding $\alpha$-Mn-type non-centrosymmetric 
sample. Upon further increasing the Mo content, in the second superconducting \tcr{region}
(for instance at $x = 0.40$), the alloy adopts a $\beta$-CrFe-type structure and exhibits a $T_c = 6.07$\,K, 
relatively lower than that of other phases. The third superconducting \tcr{region}
has its onset at $x \gtrsim 0.5$, with all the samples showing a bcc-W-type 
crystal structure and the highest $T_c$ reaching 12.4\,K at $x = 0.60$. 
In \tcr{all the regions}, specific-heat data evidence a fully-gapped 
superconducting state with a single gap.
%whereas in the second phase, \tcr{as reported in Ref.~\onlinecite{Sundar2015}}, 
%multiband superconductivity was revealed by both specific-heat and 
%magnetization measurements. For examples,  
%for the $x = 0.60$ and 0.75 cases, besides the large SC gaps of 2.13 and 
%2.08\,$k_\mathrm{B}T_c$, additional small SC gaps of 0.66 and 
%1.01\,$k_\mathrm{B}T_c$ appear. 
Compared  with the pure Re and Mo superconductors, as well as with the 
ideal BCS case (for which $\Delta_{0}/k_\mathrm{B}T_{c} = 1.76$), 
Re$_{1-x}$Mo$_x$ alloys with $0.12 \le x \le 0.75$ exhibit larger 
superconducting gap values, thus indicating moderately coupled 
superconducting pairs in the latter.

%
%==== figure =============================%
%=== end figure ==========================%
%

Figure~\ref{fig:phasediagram}(b) shows the calculated band-structure 
density of states (DOS) $N_\mathrm{band}(\epsilon_\mathrm{F}$) and 
the electron-phonon coupling 
constant $\lambda_\mathrm{ep}$ versus the Re/Mo concentration (see 
Sec.~\ref{ssec:Cp} for details). Interestingly, the $\lambda_\mathrm{ep}$ follows closely the superconducting phase diagram shown in Fig.~\ref{fig:phasediagram}(a) and
assumes its maximum values in correspondence with the highest $T_c$ 
in both the bcc-W- and hcp-Mg-phases. Compared to the pure Re or Mo cases, 
the enhanced $\lambda_\mathrm{ep}$ again indicates a moderately coupled 
superconductivity in Re$_{1-x}$Mo$_x$ alloys.  
The enhanced DOS in Re$_{1-x}$Mo$_x$, on either the Re- or Mo-rich side, 
may be ascribed %attributed 
to the filling of unoccupied $d$ bands by additional Mo or Re $d$ electrons. 
In any case, the different DOS values in Re$_{1-x}$Mo$_x$ alloys seem %to be 
closely related to their crystal structures. 

Finally, we note that recent $\mu$SR studies on non-centro\-sym\-met\-ric 
Re$T$ alloys indicate that TRS is consistently broken in their 
superconducting states. Since the TRSB occurrence in Re$T$ is independent 
of the particular transition metal, this points at the key role played by Re. 
To further confirm such conclusion, Re-containing samples with the same 
stoichiometry but with different crystal structures are ideal candidates. 
Considering their %very 
\tcb{rich} structural and electronic properties for 
different Re/Mo concentrations, here we could show that Re$_{1-x}$Mo$_x$ 
alloys indeed represent such candidates. For example, for $x = 0.23$, 
samples with either non-centrosymmetric $\alpha$-Mn-type or 
centrosymmetric hcp-Mg-type (the same as pure Re) crystal structures can 
be prepared. Our results demonstrate an intimate relationship between 
electronic- and crystal structure in Re$_{1-x}$Mo$_x$ alloys, which 
puts them forward as excellent \tcb{systems} %candidates % TOO many repetitions of "candidate" 
for studying the interplay among the various symmetries in the superconducting state.  

\vspace{7pt}
\section{\label{ssec:Sum}Conclusion}
To summarize, by combining arc melting and annealing processes, we managed 
to synthesize 
Re$_{1-x}$Mo$_x$ binary alloys in a wide range of solid 
solutions. XRD patterns are consistent with four different solid phases, 
including hcp-Mg ($P6_3/mmc$), $\alpha$-Mn($I\overline{4}3m$), 
$\beta$-CrFe($P4_2/mnm$), and bcc-W ($Im\overline{3}m$), of which 
$\alpha$-Mn and $\beta$-CrFe have been among the most difficult to obtain to date. 
As established by electrical-resistivity, magnetic-susceptibility, and 
heat-capacity measurements, across the full range of solid solutions,   
Re$_{1-x}$Mo$_x$ exhibits three superconducting \tcr{regions}, with the highest 
$T_c$ reaching 9.43 and 12.4\,K in the hcp-Mg- and $\beta$-CrFe-phases, 
respectively. The low-temperature electronic specific heat evidences a 
fully-gapped superconducting state, whose enhanced gap magnitude and 
specific-heat discontinuity suggest a moderately strong pairing,  
the latter being consistent with an enhanced electron-phonon coupling constant 
$\lambda_\mathrm{ep}$. The close correlation between $\lambda_\mathrm{ep}$, 
DOS, superconductivity, and crystal structure indicates 
that Re$_{1-x}$Mo$_x$ alloys represent a very interesting
system for studying the effects of
structural symmetry on the electronic properties. The superconducting 
phase diagram reported here paves the way to future microscopic 
investigations, including the use of local-probe techniques such 
as NMR and $\mu$SR. Forthcoming zero-field $\mu$SR experiments will 
be essential to clarify the possible influence of structure on 
the occurrence of TRSB in the Re$_{1-x}$Mo$_x$ family.

\begin{acknowledgments} 
The authors thank S. Ghosh, J. Quintanilla, P. Puphal, and I. Eremin for fruitful discussions.
This work was supported by the Schwei\-ze\-rische Na\-ti\-o\-nal\-fonds zur F\"{o}r\-de\-rung der Wis\-sen\-schaft\-lich\-en For\-schung, SNF (Grants no.\ 200021-169455 and 206021-139082).
\end{acknowledgments}

%\begin{footnotesize}
\bibliography{ReMo_bib}
%\end{footnotesize}

\end{document}